\documentclass[prl,aps,twocolumn,reprint,superscriptaddress,showpacs]{revtex4}

\tighten
\renewcommand{\narrowtext} 
{\begin{multicols}{2}\global\columnwidth20.5pc} 

\usepackage{graphicx}  
\usepackage{dcolumn}   
\usepackage{bm}        
\usepackage{amssymb}   
\usepackage{amsmath}   
\usepackage{microtype}   
\usepackage[breaklinks]{hyperref} 
\hypersetup{
  colorlinks   = true, 
  urlcolor     = darkblue, 
  linkcolor    = darkblue, 
  citecolor    = darkblue 
}

\newcommand{\be}{\begin{equation}}
\newcommand{\ee}{\end{equation}}
\newcommand{\bea}{\begin{eqnarray}}
\newcommand{\eea}{\end{eqnarray}}

\newcommand{\br}{{\bf r}}

\newcommand{\ci}{\frak{i}}

\usepackage{wasysym}
\usepackage{graphicx}
\usepackage{bm}

\newcommand{\ColorOnline}{(Color online) }
\usepackage{ulem}
\usepackage{color}
\definecolor{darkred}{rgb}{0.6,0,0}
\definecolor{darkblue}{rgb}{0.0,0,0.6}
\definecolor{red}{rgb}{1,0,0}

\newcommand{\calF}{{\cal F}}
\newcommand{\nStar}{{n^{*}}}
\newcommand{\tl}{{\tau^{*}_{\nAB}}}
\newcommand{\tll}{{\tau_{\nAB}}}

\newcommand{\nA}{n_\text{A}}
\newcommand{\nB}{n_\text{B}}
\newcommand{\nAB}{\bar n}
\newcommand{\nv}{n_{\text{vac}}}

\newcommand{\mfD}{\mathfrak{D}}

\newcommand{\mfx}{\mathfrak{x}}
\newcommand{\mfy}{\mathfrak{y}}

\usepackage{multirow}
\usepackage{textcomp}
\usepackage{tikz}
\usepackage{bigdelim}

\usetikzlibrary{shapes,external,decorations.pathreplacing,decorations.markings,calc,arrows,backgrounds}

\DeclareGraphicsExtensions{.jpg,.png,.pdf,.eps}

\newcommand{\midarrow}{\tikz \draw[-triangle 90] (0,0) -- +(.1,0);}

\newcommand{\tmatrix}{\begin{tikzpicture}[implies/.style={double,double equal sign distance,-implies},scale=0.8]
		\node [circle,fill,draw,inner sep=2pt] (1) at (0, -0) {};
		\node [cross out,draw] (2)  at (0, 2) {};
		\draw [line width=3.5,style=dashed] (1) to (2);
\end{tikzpicture}}

\newcommand{\zerothorder}{\begin{tikzpicture}[implies/.style={double,double equal sign distance,-implies},scale=0.8]
		\node [circle,fill,draw,inner sep=2pt] (1) at (0, -0) {};
		\node [cross out,draw] (2)  at (0, 2) {};
		\draw [style=dashed] (1) to (2);
\end{tikzpicture}}
\newcommand{\firstorder}{\begin{tikzpicture}[implies/.style={double,double equal sign distance,-implies},scale=0.8]
		\node [circle,fill,draw,inner sep=2pt] (1) at (0, -0) {};
		\node [cross out,draw] (2)  at (0.75, 2) {};
		\node [circle,fill,draw,inner sep=2pt] (3) at (1.5, -0) {};				
		\draw [style=dashed] (1) to (2);
		\draw [style=dashed] (3) to (2);			
	\begin{scope}[ every node/.style={sloped,allow upside down}]
				\draw (1)-- node{\midarrow}(3);\end{scope}
\end{tikzpicture}}
\newcommand{\secondorder}[0]{\begin{tikzpicture}[implies/.style={double,double equal sign distance,-implies},scale=0.8]
		\node [circle,fill,draw,inner sep=2pt] (1) at (0, -0) {};
		\node [cross out,draw] (2)  at (1.5, 2) {};
		\node [circle,fill,draw,inner sep=2pt] (3) at (1.5, -0) {};				
		\node [circle,fill,draw,inner sep=2pt] (6) at (3, -0) {};
		\draw [style=dashed] (1) to (2);
		\draw [style=dashed] (3) to (2);	
		\draw [style=dashed] (6) to (2);			
	\begin{scope}[ every node/.style={sloped,allow upside down}]
				\draw (1)-- node{\midarrow}(3);
				\draw (3)-- node{\midarrow}(6);							
  \end{scope}
\end{tikzpicture}}

\begin{document} 
\preprint{p8.grapheneCompensatedVacancies-DoS}

\title{Density of states in graphene with vacancies: midgap power law and frozen multifractality}

\author{V. H\"afner}
\affiliation{ Institute of Nanotechnology,
 Karlsruhe Institute of Technology, Campus North, D-76344
  Karlsruhe, Germany}
\affiliation{Institut f\"ur Theorie der Kondensierten Materie,
 Karlsruhe Institute of Technology, Campus South, D-76128 Karlsruhe, Germany}
\author{J. Schindler}
\affiliation{ Institute of Nanotechnology,
 Karlsruhe Institute of Technology, Campus North, D-76344
  Karlsruhe, Germany}
\affiliation{Institut f\"ur Theorie der Kondensierten Materie,
 Karlsruhe Institute of Technology, Campus South, D-76128 Karlsruhe, Germany}
\author{N. Weik}
\affiliation{ Institute of Nanotechnology,
 Karlsruhe Institute of Technology, Campus North, D-76344
  Karlsruhe, Germany}
\affiliation{Institut f\"ur Theorie der Kondensierten Materie,
 Karlsruhe Institute of Technology, Campus South, D-76128 Karlsruhe, Germany}
\author{T. Mayer}
\affiliation{ Institute of Nanotechnology,
 Karlsruhe Institute of Technology, Campus North, D-76344
  Karlsruhe, Germany}
\affiliation{Institut f\"ur Theorie der Kondensierten Materie,
 Karlsruhe Institute of Technology, Campus South, D-76128 Karlsruhe, Germany}
\author{S. Balakrishnan}
\affiliation{Department of Physics, Indian Institute of Technology Madras,
Chennai 600036, India}
\author{R. Narayanan}
\affiliation{Department of Physics, Indian Institute of Technology Madras,
Chennai 600036, India}
\author{S. Bera}
\affiliation{ Institute N\'eel and Universit\'e Grenoble Alpes, F-38042 Grenoble, France}
\author{F. Evers}
\affiliation{ Institute of Nanotechnology,
 Karlsruhe Institute of Technology, Campus North, D-76344
  Karlsruhe, Germany}
\affiliation{Institut f\"ur Theorie der Kondensierten Materie,
 Karlsruhe Institute of Technology, Campus South, D-76128 Karlsruhe, Germany}
\affiliation{Center of Functional Nanostructures, 
 Karlsruhe Institute of Technology, Campus South, 
  D-76131 Karlsruhe, Germany}

\date{\today}

\pacs{73.22.Pr, 72.80.Vp, 71.23.-k} 
\keywords{ }
\begin{abstract}
The density of states (DoS), $\varrho(E)$,  
of graphene is investigated numerically and within the self-consistent T-matrix approximation 
(SCTMA) in the presence of vacancies within the tight binding model.
 The focus is on compensated
disorder,  where the concentration of vacancies, 
$\nA$ and $\nB$, in both sub-lattices is the same. Formally, this model belongs to the 
chiral symmetry class BDI. The prediction of the non-linear sigma-model 
for this class is a Gade-type singularity $\varrho(E) \sim |E|^{-1}\exp(-|\log(E)|^{-1/x})$. 
Our numerical data is compatible with this result 
in a preasymptotic regime that gives way, however, at even lower energies 
to $\varrho(E)\sim E^{-1}|\log(E)|^{-\mfx}$, $1\leq \mfx < 2$. We take this finding as an evidence that similar 
to the case of dirty d-wave superconductors, also generic bipartite random hopping models
may exhibit unconventional (strong-coupling) fixed points for certain kinds of 
randomly placed scatterers if these are strong enough. 
Our research suggests that graphene with (effective) vacancy disorder is a 
physical representative of such systems. 
\end{abstract}
\maketitle

Graphene is a hot topic in material sciences and 
condensed matter physics~\cite{CN09}.
The material is interesting 
its electronic structure hosts 
two Dirac-cones.  
Since only the $\pi_z$-orbitals make significant 
contributions to the relativistic sectors of the 
band-structure, 
a tight-binding description of the material is 
frequently employed that keeps a single orbital per 
carbon atom and only nearest-neighbor hopping. 
Within this description it is easy to see that 
disorder introduced by a random distribution of vacancies 
has nontrivial effects. For instance, 
it is well known that a single impurity populates 
a mid-gap state which is power-law localized
\cite{PEREI06,PEREI08}.
With a finite concentration of vacancies a rich 
plethora of new phenomena emerges. 
One distinguishes the ``compensated'' case,  
-- same concentration of vacancies 
in each sub-lattice, $\nAB{=}\nA{=}\nB$ 
-- from the uncompensated case, $\nA{>}\nB$. 
In the latter case, one expects that the DoS exhibits 
a (pseudo-) gap, while for compensated disorder 
a sharp peak is observed \cite{CN09}. 
Most studies focus on the balanced case 
at concentrations well below the percolation threshold, 
$\nAB\lesssim 30\%$. 
At present only very few aspects have been investigated 
in detail, despite the importance 
of the DoS for transport and optical properties of the 
functionalized material~\cite{ABA10,wehling10}.

Graphene with vacancies represents a bipartite random hopping 
system with time reversal and spin rotational invariance. Following 
the Zirnbauer-Altland classification of disordered metals it belongs to 
symmetry class BDI, \cite{ALT97,EVE08}. In the presence of weak bond disorder, 
a description based on the non-linear $\sigma$-model 
predicts for the density of states
\be
\label{e1}
\ln |E \varrho(E)| \sim  -|\ln (E/\mfD)|^{1/x}, \qquad |E|\lesssim \mfD
\ee
where $\mfD(\nAB)$ denotes a microscopic energy scale.~\cite{EVE08}
The exponent $1/x$ reflects a peculiar feature of the RG-flow 
found by Gade and Wegner in a perturbative renormalization group (RG) 
study~\cite{gade91,gade93}. Their analysis shows 
that the energy flow with the RG-scale $L$ is  
$
|\ln \epsilon| \propto z(L) |\ln L|. 
$
Unlike the case with conventional critical behavior,
the dynamical exponent $z$ is not a constant here but rather $z(L)\propto \ln L$, 
so $|\ln \epsilon|\propto |\ln L|^2$ and correspondingly  
an exponent $x{=}2$ was obtained 
\footnote{The result was confirmed by Guruswamy et al.~\cite{guruswamy99} 
in their analysis of a bipartite $\pi$-flux model that also belongs to class BDI. 
Specifically, this study shows that generic representatives of BDI exhibit a 
running coupling, 
$g_A \sim \ln L$. 
Within the $\pi-$flux model $g_A$ has the interpretation 
of a coupling to a real random gauge-field. Since one has  $z\approx 1 + 2 g_A$, 
the previous 
conclusion $z(L)\sim \ln L$ is a consequence of ``runaway flow''. }.

Later it was argued that the logarithmically growing exponent $z$ 
is an indication of ``freezing''~\cite{motrunich02,mudry03}. 
Freezing sets in when disorder has become so strong
that critical wavefunctions concentrate in rare regions of the sample with 
very weak, power law tails leaking out of these ``optimal'' domains. 
In such situations, observables that  derive from moments 
of wavefunction amplitudes higher than the first one 
cease to be sensitive to the sample geometry, 
so that their ``flow'' with the system size is ``frozen''. 
Freezing implies that at $z{\geq} 3$ rare events dominate 
the energy-scaling and a new dependency 
$z\approx 4\sqrt{\ln L}{-}1$ sets in~\cite{EVE08}. 
As a consequence, the Gade-exponent $x{=}2$ gives way to $x{=}3/2$
and the zero-energy singularity becomes slightly {\it weaker}
in the frozen limit. 

A strong increase of the DoS 
near zero energy has been observed 
in several numerical works~\cite{PEREI08,YUA10,stauber08,wu08,wehling10}, 
but a quantitative check of the prediction, Eq. \eqref{e1},  
is still missing. 
Here, we present such an analysis. 
We confirm the existence of a parametrically wide energy window 
where $\varrho(E)$ indeed follows Eq. \eqref{e1}. 
However, at ultra-low energies, 
Eq. \eqref{e1} is not valid. Instead, 
the DoS crosses over to new behavior with a  
significantly {\it stronger} singularity,
$1/(E|\log(E)|^{{\mfx}})$, with $2>\mfx\geq 1$. 

\paragraph*{Model and Methods (MaM): SCTMA.}
For the SCTMA we adopt the formalism developed 
in an earlier work 
and use it here to calculate the DoS~\cite{ABA10,ostrovsky06}.
\paragraph*{MaM: Stochastic time evolution.} 
The SCTMA results are then compared
against numerical simulation data for $\varrho(E)$
as obtained from a tight-binding Hamiltonian of the honeycomb lattice
$\hat H = -t \sum_{<ij>} c^\dagger_{i}c_{j}$ where as usual $<ij>$ 
indicates nearest neighbor hopping. The disorder average is performed 
at vacancy concentration $\nAB$ fixed and the same for both sublattices. 
We employ a numerical technique similar to Ref.~\cite{YUA10} 
exploiting 
$\varrho(E){=} \int_{-\infty}^\infty d\tau \varrho(\tau) e^{\ci E\tau}$ 
with the exact stochastic representation 
\be
\label{e2}
\varrho(\tau) = \frac{1}{2\pi}\overline{\left\{ \langle\phi|\exp(-\ci \hat H \tau)|\phi\rangle \right\}_\text{in}}. 
\ee
Here, $|\phi\rangle$ represents a random initial state and $\{\ldots \}_\text{in}$ denotes 
an ensemble average of such states. For the evaluation of the matrix element we employ a 
standard Krylov-subspace approach, with a conservative choice of the 
width of the integration steps, typically $dt{=}0.01$ (units: $t^{-1}$), 
and an observation window of $10^6$ steps corresponding to a time
$T_\text{obs}=10^4$~\footnote{At our system sizes, $L=2048,4096$, we did not
  observe a significant
effect of $\phi-$averaging on $\varrho(E)$ due to self-averaging. 
If not specified otherwise, averaging was over four initial
states~\cite{hafner2011,schindler2012}. A convergence test justifying the 
choice of $dt$ is given in the supplementary material.}.
In order not to lose correlations due to methodological artifacts
over the observation time, 
the time increment $dt$ should become significantly 
smaller with growing $T_\text{obs}$. As it turns out, 
this makes the stochastic time evolution numerically highly demanding 
at ultra-low energies. 

\paragraph{MaM: Generalized multifractal analysis~(GMA).} 
In order to explore $\varrho(E)$ at ultra-low energies, 
we first calculate the localization length, $\xi(E)$, 
with spectral methods which in turn is closely related to $\varrho(E)$:  
If one assumes that a localization volume
$\xi^2$ has typically one state with lowest 
energy $E_\xi$ one has: $|E_\xi|\xi^2\varrho(E_\xi) = {\cal O}(1)$. Hence
$\xi(E) \approx |E \cdot \varrho(E)|^{-1/2}.$  
The expression is familiar from the standard weak coupling analysis~\cite{EVE08}.
A complication arises because the same analysis 
predicts the form Eq. \eqref{e1} for the DoS that turns out 
incompatible with our numerical data - as we already mentioned. 
Hence, a more general form 
$|E_\xi|\xi^2\varrho(E_\xi) = r(\ell/\xi)$ should be considered
($\ell$: a microscopic length). 
Partially inspired by most recent analytical work \cite{ostrovsky14},
we argue in the supplementary material
that a reasonable assumption would be 
$r(X)\approx 1/X^{\mfy}$ (with $\mfy=1$) at energies not too low, 
so that 
\be
\label{e3}
\xi(E) \approx \left| E \cdot \varrho(E)\right|^{-1/(2+\mfy)}.   
\ee 

Spectral methods allow us to extract the localization length even at very low energies 
and therefore can provide information about the DoS as well. 
Similar to Ref. \cite{rodriguez10}, we 
employ a generalized multifractal analysis (GMA) for this purpose.
 It is motivated in the present context 
from the fact that multifractality at the 
Dirac-point is a topic of interest {\it per se}. 
The central observable is  the inverse participation 
ratio (IPR), 
$
P_q (E) =  \int_{L^2} d\br |\psi_{m}(\br)|^{2q}, 
$
(For numerical efficiency, we average over a small number of 
states with energies $\epsilon_m$ inside an interval about $E$. 
In addition, we also perform a disorder average - at fixed $\nAB$ -  
that we indicate via $\overline{P_q}$.) 
To address the localization length, $\xi(E)$, 
one works at finite energies $|E|>0$ in the vicinity of the critical 
fixed point where a scaling Ansatz 
\be
\label{e5}
\overline{P_q} = L^{-\tau_q} \calF_q(L/\xi(E)) 
\ee
applies. We will extract $\xi(E)$ by scaling our numerical data 
according to this relation.  The wavefunction data 
has been obtained  in a well documented way 
(e.g. Ref.~\cite{evers01,subramaniam06}) 
employing standard sparse matrix routines~\cite{arpack}. 


\begin{figure}[bp!]
\includegraphics[width=0.99\linewidth]{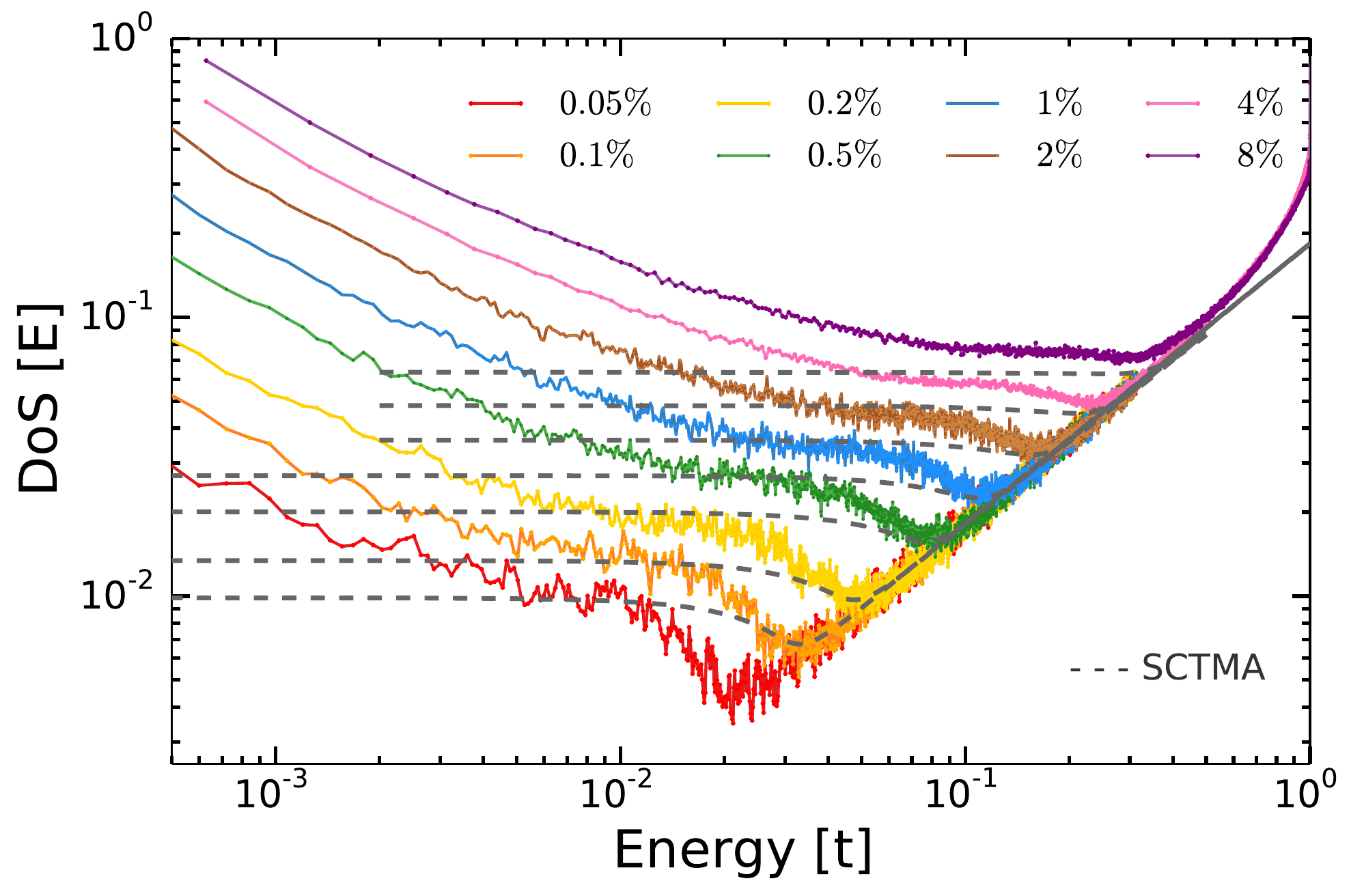}
\caption{\ColorOnline Density of states of graphene with $\nAB{=}0.1-8\%$ 
vacancies in either sublattice. Comparison of SCTMA and 
tight-binding simulation. 
}
\label{f1}
\end{figure}

\paragraph{Results: SCTMA.} The DoS as obtained from the self-consistency cycle of the SCTMA is shown in 
Fig.~\ref{f1} with dashed lines.  
In the limit of large and low energies we recover the expected qualitative behavior: 
If the energy exceeds a characteristic scale set by 
$
\Delta(\nAB) {=} v_\text{F}\sqrt{\pi \nAB/\ln(\nStar/\nAB)   }
$, with $\nStar{=}W^2/\pi v_\text{F}^2$, and $W$ a high-energy cut-off, 
\cite{ostrovsky06}, 
the DoS essentially remains unaffected by the impurities thus retaining the characteristic 
linear form reminiscent of clean graphene at high energies. 
(Our data suggests $n^*{\approx} 1$.) However, in the low-energy limit,
$E{\ll}\Delta(\nAB)$, the Dirac-singularity broadens 
and one obtains a constant value for the DoS. 

As seen in the expression for the characteristic energy scale $\Delta(\nAB)$, 
the SCTMA provides a  
logarithmic renormalization of the naive scale  
$\sqrt{\pi \nAB v_\text{F}^2}$ that follows 
from dimensional analysis. In similar vein, 
in the limit $E\to 0$, our data suggests that the saturation 
value of the DoS picks up similar logarithmic corrections, 
$\varrho^\text{SCTMA}(0)\sim \Delta(\nAB) \ln(\nStar/\nAB)$. 
Furthermore, this logarithmic dressing 
leads to the minimum in the DoS as seen in Fig.~\ref{f1}~\footnote{Namely, the high energy trace follows the 
unperturbed behavior $\varrho(E) = |E|/\pi\sqrt{3}$ all the way down to $\Delta(\nAB)$. 
At the departure point into the low-energy region,  
$\varrho(\Delta(\nAB))$  exhibits a DoS already below 
the limit $\varrho^\text{SCTMA}(0)$ and so the DoS increases again.}. 



\begin{figure}[tbp!]
\includegraphics[width=0.975\columnwidth]{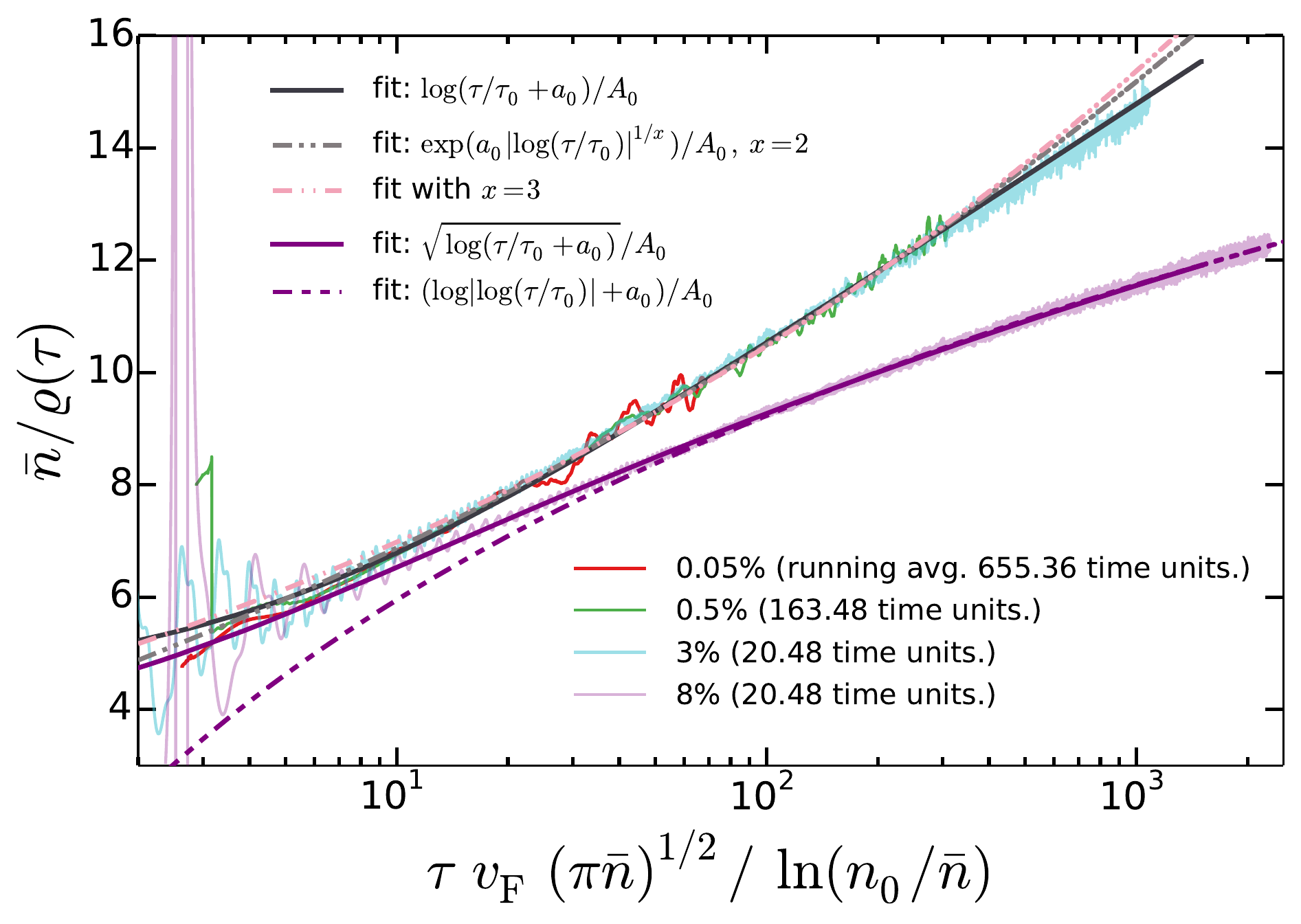}
\caption{\ColorOnline Data collapse of inverse time-series on a master curve, 
consistent with the Wegner-Gade scaling Eq. \eqref{e1}. 
Solid lines represent according fits: 
$\exp(\mathfrak{a}_0|\ln(\tau/\tau_0)|^{1/x})/\mathfrak{A}_0$
($x{=}2: (\tau_0,\mathfrak{a}_0,\mathfrak{A}_0)=(0.0169, 1.004, 1.838)$; 
$x{=}3/2: (0.00013, 0.614, 3.2)$)
(Times $\tau_0$ are measured in units 
$(\pi\nAB v_\text{F}^2)^{1/2}/\ln(n_0/\nAB)$.
We fix $n_0\approx 2.08$ by collapsing onto the master curve.) 
Collapse restricts 
to a pre-asymptotic time window, $1\ll \mfD\tau \ll \mfD\tl$ 
(displayed: $\nAB=0.05\%, 0.5\%, 3\%$).  
After a crossover to ultra-long times, 
$\tau\gg \tl$, the increase of $\nAB/\varrho(\tau)$ 
 is sublinear; an example numerically accessible in this time regime 
is the $\nAB=8\%$-trace. 
We fit 
$\sqrt{\ln (\tau/\tau_0+\mathfrak{a}_0)}/\mathfrak{A}_0$ ($(\tau_n,\mathfrak{a}_n,\mathfrak{A}_n)=(1.671,1.74,0.219)$)
motivated by Ref. \cite{ostrovsky14}. 
Dashed lines guide the eye indicating alternative fittings:  
$\ln(\tau)$ 
and  
$\ln\ln(\tau)$. 
The fluctuations in the raw data reflect the stochastic nature of the methodology. 
 }
\label{f2}
\end{figure}

\paragraph{Results, Tb-simulation:~energy.}
Since the SCTMA ignores multiple 
scattering at two- (or more) impurity configurations, quantum-interference processes are absent. 
Hence, within the SCTMA one does not expect any indication of the $E^{-1}$ singularity 
predicted in Eq. (\ref{f1}). To investigate this,  
we resort to a numerical simulation of the DoS in the lattice model.
As one might have suspected, 
the characteristic minimum in the DoS obtained within the SCTMA
is also seen in the lattice simulation 
Fig.~\ref{f1} and turns out to be even more pronounced there. 
Quantum interference becomes important at energies
below a scale  $\mfD(\nAB)$ where it gradually enhances the 
(upturning) curvature.

\paragraph{Results, Tb-simulation:~time.}
At lowest energies the Fourier-transformation~(FT) exhibits a sensitivity to the 
window of integration times. Even though artifacts are generally weak, 
for the present purpose we will work in the time representation 
and eliminate (residual) high-frequency contributions to 
$\rho(\tau)$ via running time averages  
(averaging windows: 20.48 - 655.36 time units); 
observation time $T_\text{obs}=10^4$. 
Fig.~\ref{f2}  displays the first out of the two key results of this 
work: at intermediate times the DoS takes a form consistent with 
Eq. \eqref{e1}
\be
\label{e6}
\varrho(\tau)\approx  {\nAB \ \mathfrak A}_0\exp\left[-\mathfrak{a}_0|\ln(\tau/\tau_0)|^{1/x}\right],  
\quad \mfD^{-1}\ll \tau \ll \tl. 
\ee
\begin{figure}[tbp!]
\includegraphics[width=0.975\columnwidth]{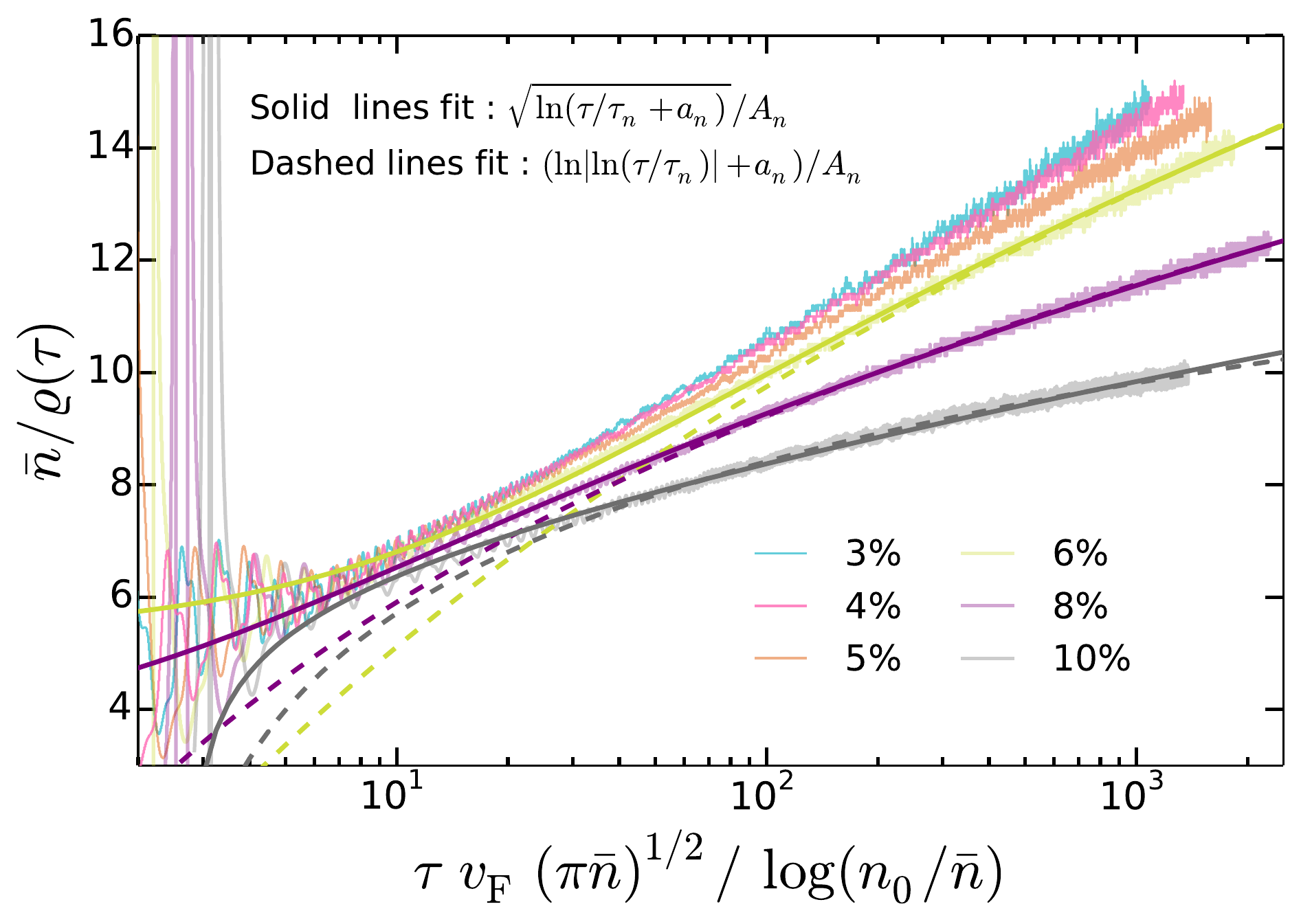}
\caption{\ColorOnline 
Evolution of $\nAB/\varrho(\tau)$ into the ultra-long time regime. 
Concentrations:
$\nAB=3\%, 4\%, 5\%, 6\%, 8\%, 10\%$.  
Fits are indicated according to
$\sqrt{\ln(\tau/\tau_{\nAB}+\mathfrak{a}_{\nAB})}/\mathfrak{A}_{\nAB}$ (solid):  
$6\%: (\tau_{\nAB},\mathfrak{a}_{\nAB},\mathfrak{A}_{\nAB})=(6.74,2.27,0.169)$;
$8\%: (1.671, 1.74, 0.219)$; 
$10\%: (0.218, -11.76, 0.295)$
and $(\ln|\ln\tau/\tau_{\nAB}|+\mathfrak{a}_{\nAB})/\mathfrak{A}_{\nAB}$ (dashed): 
$8\%: (\tau_{\nAB},\mathfrak{a}_{\nAB},\mathfrak{A}_{\nAB})=(0.109,-0.77,0.124)$;
$10\%: (2.034, 1.425, 0.33)$. }
\label{f3}
\end{figure}
The crossover scale $\tl$ is very rapidly decreasing if $\nAB$ grows from 
3\% to 10\%. As a consequence, the onset of the ultra-long time regime 
can be investigated with the time propagation method. 
As shown in Fig.~\ref{f3}, at times exceeding $\tl$ the decay of $\rho(\tau)$ 
is much slower even than $1/\ln(\tau)$. The accessible time window is too small in order 
to reliably discriminate possible cases, $1\leq \mfx < 2$, 
\be
\label{e7}
\varrho(\tau) = \nAB \mathfrak{A}_{\nAB}|\ln(\tau/\tau_{\nAB}+\mathfrak{a}_{\nAB})|^{-\mfx+1}, 
\qquad \tl\ll \tau.  
\ee
(Even $\mfx{\to} 1$, i.e. $\varrho(\tau){=}\nAB\mathfrak{A}_n / ( \ln( \ln(\tau/\tll)){+}\mathfrak{a}_{\nAB})$, 
would not be incompatible with the data~(see Figures~\ref{f2} and \ref{f3}).)  
What can safely be concluded at this point is that at very low energies
$|E|\cdot \varrho(E)\propto 1/\ln(|E|)^{{\mfx}}$, $1\leq {\mfx}<2$ at 
variance with Eq. \eqref{e1}.

\begin{figure}[tp]
\includegraphics[width=0.45\columnwidth]{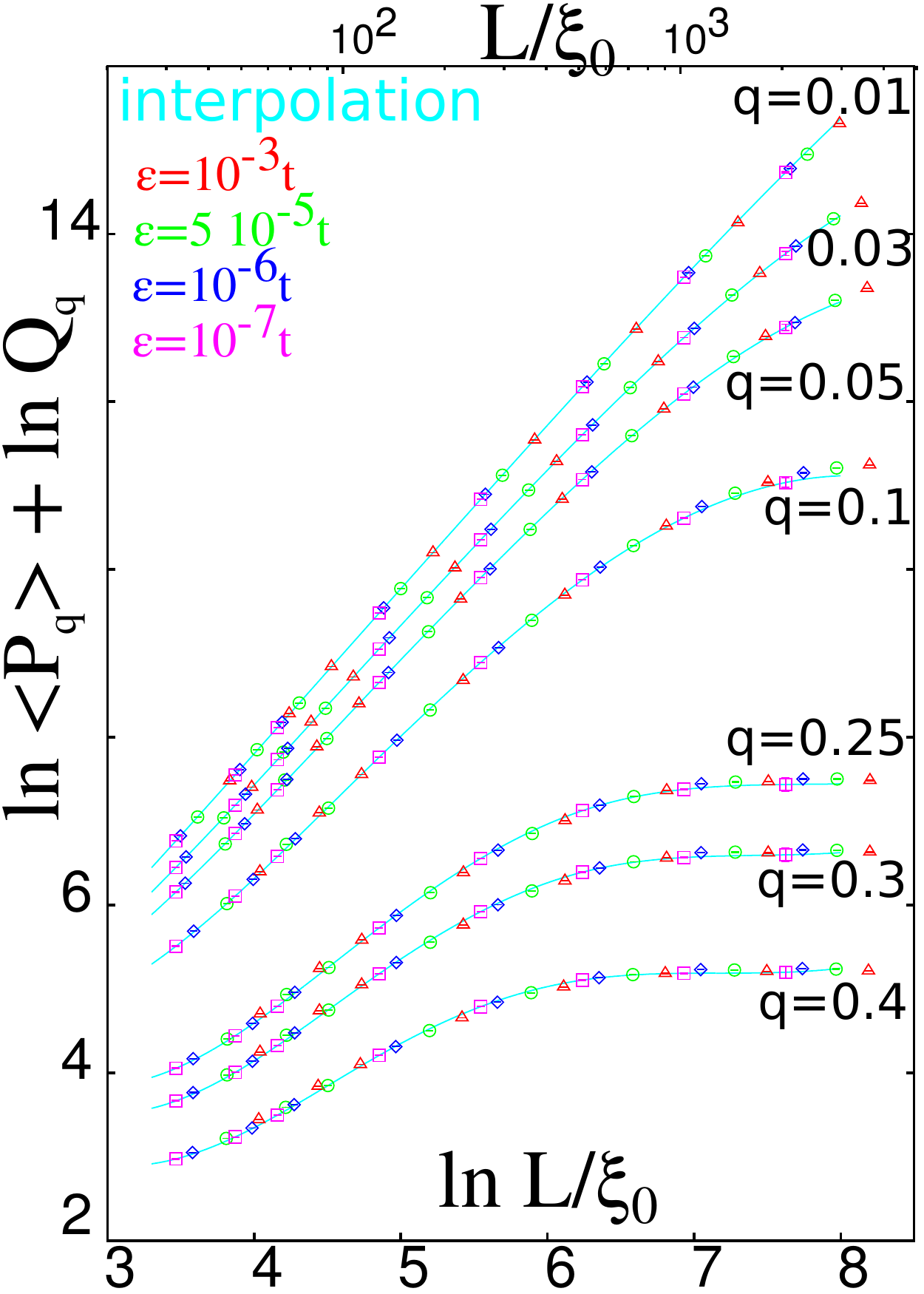}
\includegraphics[width=0.45\columnwidth]{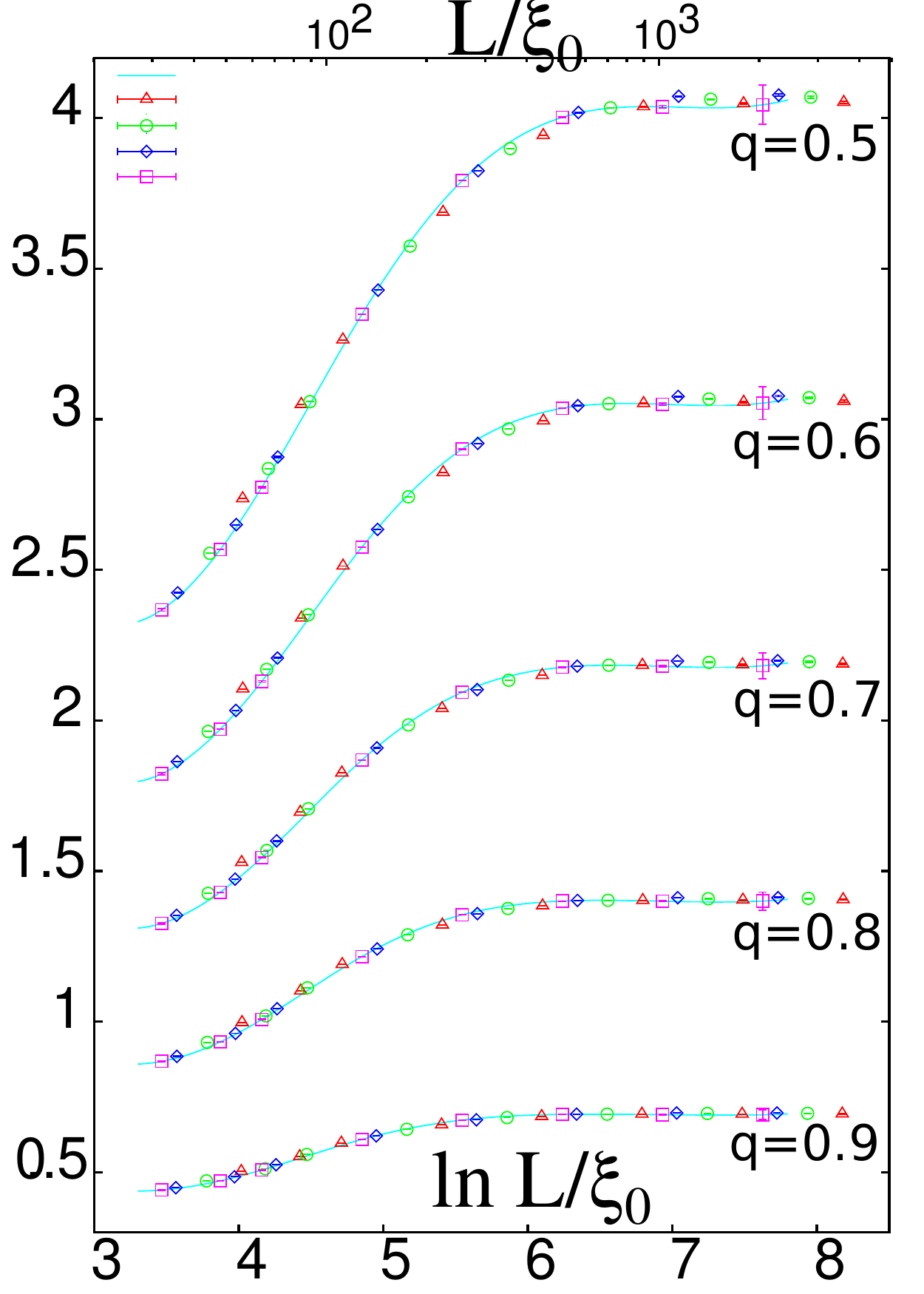}
\includegraphics[width=0.95\columnwidth]{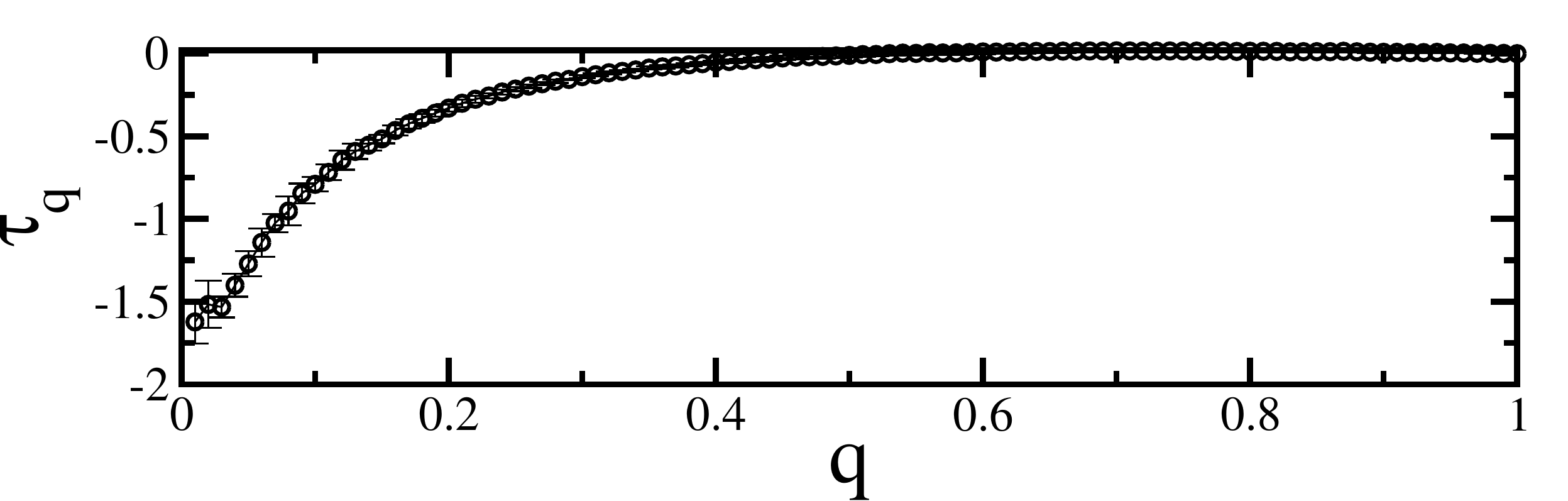}
\caption{\ColorOnline Top panels: master curves for different $q$-values 
as obtained after re-scaling of $x,y-$axes 
with energy-depending scale factors 
$\xi(E)$ ($x-$axis, depicted in Fig.~\ref{f3}) and 
$Q_q=\xi(E)^{\tau_q}$ ($y$-axis, not shown). 
Parameters: $\nAB=4\%$, $L= 64, 128, 256, 512, 1024, 2048$,  
$\epsilon=10^{-3},5\times10^{-5}, 10^{-6}, 10^{-7}$. 
IPR-distribution functions are given in the supplementary material.
Bottom panel: multifractal spectrum as estimated from fitting 
to $Q_q$. It displays frozen multifractality. 
}
\label{f5}
\end{figure}

\paragraph{Results: Generalized multifractal analysis}  
We have calculated the IPR near four different energies covering the 
range $10^{-3}t-10^{-7}t$. 
The resulting master curve ${\cal F}$ defined in Eq. 
\eqref{e5} is displayed in Fig.~\ref{f5}. 
In the regime of large system sizes $L\gg \xi(E)$ 
all curves exhibit a plateau indicating that 
the IPR is independent of the growing system size: 
we observe the insulating behavior expected for the AI-class 
that eventually governs all energies except $E{=}0$. 
At smaller $L/\xi(E)$-values a  power-law regime 
begins to develop that governs intermediate system sizes but is cut off
at smallest values $L\ll \xi(E)$ where the slope begins to decrease again. 
This peculiar feature foreshadows 
the behavior at the critical fixed point. We believe that it 
indicates the existence of a second plateau in the limit $\xi\to\infty$ 
that exists at $q_c\leq q < 1$ and 
that is not yet fully developed in our data. 
This plateau is manifestation that certain moments, $q>q_c$, also 
of the critical wavefunctions become insensitive of the system size 
growth and are (in this sense) ``frozen''. 

\begin{figure}[tb]
\includegraphics[width=0.99\columnwidth]{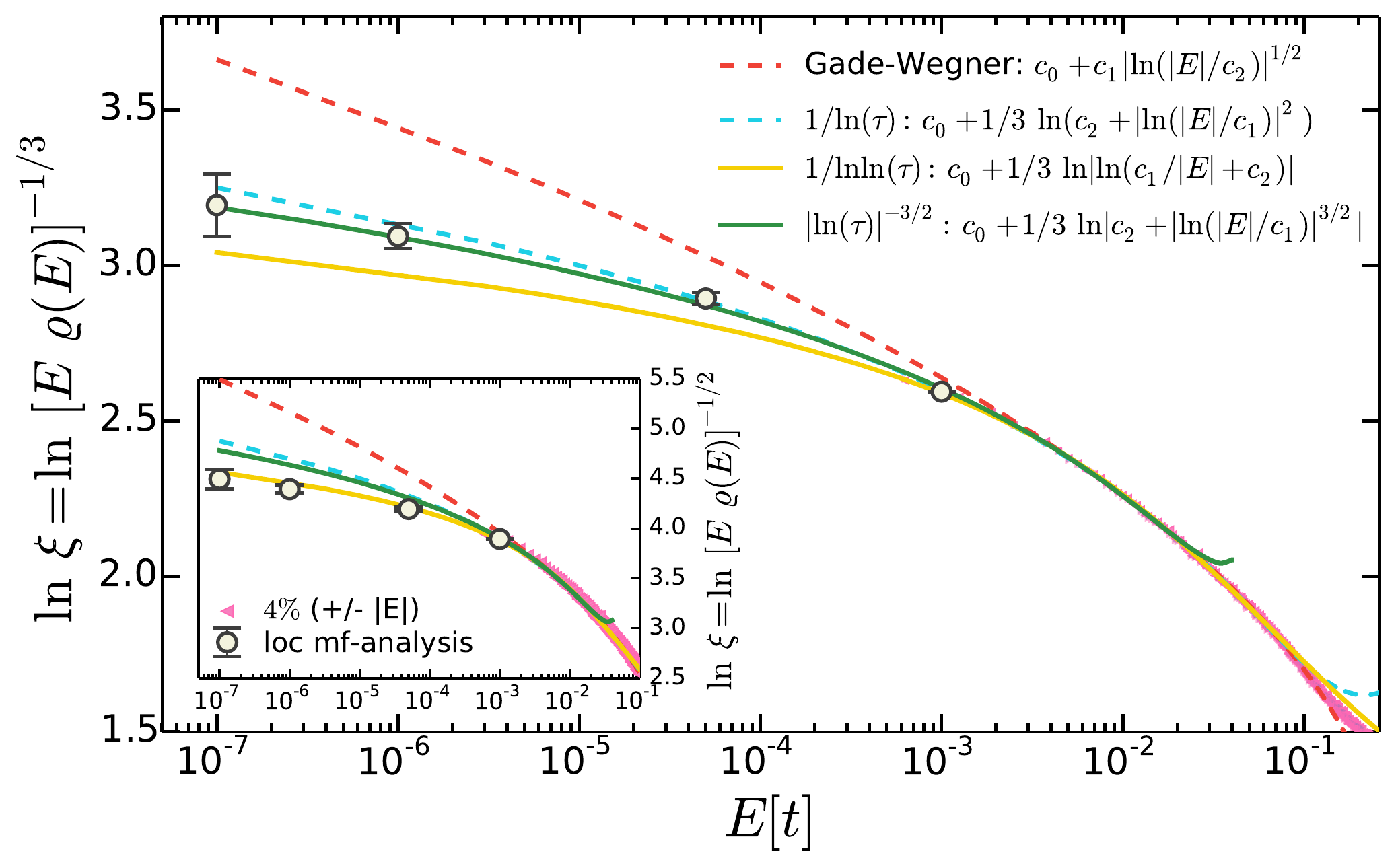}
\caption{\ColorOnline Localization length as obtained from the 
DoS (data Fig.~\ref{f1}. solid) and from the GMF-analysis (symbols, converted 
assuming $\mfy{=}1$.). Also shown 
are three fits to the high energy regime (see inset) 
that interpolate into the 
low energy section via FT of 
Eq. \eqref{e6} (blue, Gade-Wegner-form, (2.192, 0.215, 1.6)) and 
Eq. \eqref{e7} (yellow, $\mfx{=}3/2$, (3.307, 0.0226, 1.0428)).
[FT of Eqs. \eqref{e1} (red trace, $\mfx{=}2$, $(c_0,c_1,c_2)=(0.1487, 1.0395, 0.285)$)
is also shown for comparison even though $\mfx{=}2$ is already 
excluded from time series analysis.] 
Only Eq. \eqref{e7} corresponding to $|E|\cdot \varrho(E)\sim 1/\ln(|E|)^{3/2}$ 
fits all regimes (with three fitting parameters). Inset: Conversion of DoS into $\xi(E)$ 
assuming $\mfy{=}0$. Comparison illustrates that  $\mfy$ indeed enters the data 
interpretation in an important way, since for $\mfy{=}0$ only $\mfx{=}1$ would provide  
an acceptable fit.  }
\label{f6}
\end{figure}

Collapsing the IPRs on the master curve, Fig.~\ref{f5}, 
delivers $\tau_q$ and $\xi(E)$ in units of
$\xi_0\equiv\xi(E_0)$ for a reference energy $E_0$. The multifractal 
spectrum $\tau_q$ is displayed in Fig. \ref{f5}, lower panel. It supports 
the freezing scenario and gives a rough estimate $q_c\lesssim 0.5$. 
The localization length is 
shown in Fig.~\ref{f6} and compared with the DoS-data converted into 
$\xi(E)$ via Eq. \eqref{e3}. 
(By matching both $\xi$-traces at $\epsilon{=}10^{-3}t$ 
we fix the GMA-scale $\xi_0$.) 
The result is satisfactory in the sense that the matching procedure 
delivers a curve that smoothly interpolates from the high-energy (SCTMA) 
into the ultralow energy regime. 
This trace summarizes our second key statement. 
Namely, a consistent fit is achieved with $\mfy{=}1$ and $\mfx=3/2$ 
over data spanning more than 5 orders of magnitude in energy. 
This result is in full agreement 
with the prediction by Ostrovsky {\it et al.} \cite{ostrovsky14}.

\paragraph{Conclusions: General implications.} 
A first important conclusion from our numerical study is that the canonical
$\sigma$-model of symmetry class BDI does not apply to to the case of graphene
with vacancies. One expects that the underlying reason is related to the fact
that vacancies in the tight-binding representation should be understood as very
strong (``unitary'') scatterers that enforce zero amplitude of the scattering
wavefunction at the scattering center.  In this sense the individual scatterer
is never weak, which is at odds with the assumption underlying the derivation of the
$\sigma-$model. 

In principle, the observation that not only the symmetry class (here BDI), but also the 
type of disorder plays a crucial in determining the low-energy behavior has been 
made before~\footnote{ For instance, symmetry class D: It allows for different lattice models
(e.g. of the Chalker-Coddington type) that exhibit very different phase diagrams depending 
on the form of the disorder. Some may 
exhibit only (topologically different) localized phases (RBIM), 
but others may also support metallic phases (Cho-Fisher model). 
For a review see Ref.~\cite{EVE08}.}. 
Of particular interest here are disordered $d-wave$ superconductors with very
strong scatterers. They belong to chiral class AIII, which is the unitary cousin
of BDI. Its $\sigma$-model also exhibits the Gade-singularity, Eq.
\eqref{e1}~\cite{altland02a}. In this context an interesting proposal deviating from the
Gade-Wegner form has been made~\cite{chamon01,pepin01} (e.g., $\varrho(E)\sim
1/|E\ln(E)^2|$. i.e. $\mfx{=}2$ in our nomenclature), but so far its status has been 
controversial~\footnote{For an overview, see e.g.
Refs.~\cite{hirschfeld02,balatsky06}}.

In a recent study~\cite{willans11}, a very similar model, the Kitaev model that 
has a representation in terms of a bipartite random hopping problem of 
Majorana fermions on a hexagonal lattice in the background of $Z_2$ fluxes has been shown to have 
a similar  singular DoS with $\mfx\approx 1.7$.  However, these results were obtained in the 
gapped phase of the model, wherein the isolated impurity states are exponentially 
localized- as opposed to a $1/r$-envelop of vacancy induced zero modes in
graphene.  Hence, the relation of this result to graphene with vacancies is
uncertain. 



\paragraph{Conclusions: Microscopic realizations and graphene.}
From the point of view of graphene research, the relevance of our results
depends on the applicability of the approximation of disorder as an ensemble of
unitary scatters.  Such are realized at least approximately, e.g., when a carbon
atom forms a chemical bond with an absorbate and therefore is taken from the
sp2- into the sp3-hybridization. Indeed, an isolated sp3-hybrid induces a state
typically of the order of 10meV away from midgap~\cite{wehling10}. The zero-mode
of the tight-binding vacancy should be understood as an approximation for such a
state. Correspondingly, we might expect that the structure of the DoS, that we
study here, could be representative for the real material on the scale of
several meV, i.e. well above  $10^{-3}t$. Hence, the intermediate energy window,
which displays the quantum interference enhanced increase of the DoS, should
still be experimentally accessible, but the ultra-low energy range might prove
difficult to reach. 

\paragraph{Acknowledgments}
We thank J. Chalker, I. Gornyi, A. Mirlin, Chr. Mudry, H. Obuse and 
P. Ostrovsky for helpful discussions. 
Especially, we express our gratitude to I. Gornyi and P. Ostrovsky 
for sharing analytical results with us prior to publication. We acknowledge the DFG (CFN
and SPP 1243) for financial support. We also thank I. Kondov and the J\"ulich
Supercomputer Center (JUROPA, project HKA12) for computational assistance and
resources. 
\bibliography{all}


\newpage

\def\theequation{S\arabic{equation}}
\def\thefigure{S\arabic{figure}}

\setcounter{figure}{0} 
\setcounter{table}{0}
\setcounter{equation}{0}
\onecolumngrid 

\vspace{1.0cm}
\begin{center}
{\bf \large Supplementary material for 
``Density of states of graphene with vacancies: midgap power law and frozen
multifractality''}
\vspace{0.5cm}

\begin{quote}
{\small
We present technical details, such as analytical derivations and convergence tests, 
and additional data together with further arguments in support of the results 
reported in the main text. 
In the first part, we derive the DoS of graphene in the presence of compensated vacancy
disorder using the SCTMA. In second part we provide details on convergence of the Krylov
propagation method. Finally in the last section preliminary evidence of
freezing transition in the IPR distribution of flow has been reported. Finally,
we provide a heuristic argument about the fluctuation effects on the local
density of states and its effect on the exponent $\mfy$. 
}
\end{quote}

\end{center}

\section{Self-consistent T-matrix approximation}
In this section we briefly recall the selfconsistent T-matrix approximation
for vacancy scattering in graphene. 
\subsection{Disorder Potential}
A vacancy is a short-range impurity to be modeled by an impurity potential
that mixes states only that within in the same sublattice as the
vacancy~\cite{SAND02,SOST06}. 
Let
\begin{equation}
\Psi=(\Psi_{A,K},\Psi_{B,K},\Psi_{B,K'},\Psi_{A,K'})^T
\end{equation}
be the four-component wave-function in $A,B$-space of the sublattices and
$K,K'$-valley space. In this representation the impurity potential of an
impurity in sublattice $A$ has the following shape~\cite{SAND02,SOST06}
\begin{equation}
   V^A_k(r)=\begin{pmatrix}
V_0&0&0&V_0e^{-2i\mathbf{k_0\cdot r}}\\
0&0&0&0\\
0&0&0&0\\
V_0e^{2i\mathbf{k_0\cdot r}}&0&0&V_0
 \end{pmatrix}\cdot e^{-i\mathbf{k\cdot r}},
 \label{eq:potential}
\end{equation}
where $\mathbf{k_0}=\mathbf{K}-\mathbf{K'}$ and $V_0$ is proportional to the
impurity potential $V$~\cite{SOST06}.  
Accordingly, the scattering potential for an impurity in sublattice $B$~\cite{SOST06} is given by 
\begin{equation}
V^B_k(r)=\begin{pmatrix}
0&0&0&0\\
0&V_0&V_0e^{-2i\mathbf{k_0\cdot r}}&0\\
0&V_0e^{2i\mathbf{k_0\cdot r}}&V_0&0\\
0&0&0&0
 \end{pmatrix}\cdot e^{-i\mathbf{k\cdot r}}.
 \label{eq:potential_B}
\end{equation}

\subsection{Selfconsistent T-Matrix Approximation}
The \textit{$T$-matrix approximation} for impurity scattering entails the neglect of all diagrams with 
crossing of impurity lines~\cite{SHU08,SALT06}. The $T$ matrix 
can be expressed as the following geometric series of diagrams~\cite{SHU08,SOST06,SALT06}
\begin{equation}
\vcenter{\hbox{\tmatrix}}\ =\ \vcenter{\hbox{ \zerothorder}}\ +\ \vcenter{\hbox{\firstorder}}\ +\ 
\vcenter{\hbox{\secondorder}}\ +\ \dots\ .
\label{eq:t-matrix}
\end{equation}
The usual diagrammatic notation is applied where crosses denote scattering off the impurity with potential $V$ 
and the propagators denote the bare Green's function~\cite{SOST06,SHU08}. 
\begin{equation}
G_0(\epsilon,k)=\frac{\epsilon+v_F\tau_3\mathbf{\sigma}\cdot \mathbf{k}}{\epsilon^2-v_F^2k^2}.
\label{eq:Greenfree}
\end{equation}

Evaluating the geometric series one obtains
\begin{equation}
T=V\sum_{n=0}^\infty(G_0V)^n=\frac{V}{1-G_0V}\overset{V\rightarrow\infty}{\sim}\frac{1}{G_0}, 
 \label{eq:T-Matrix}
\end{equation}
and the $T$-matrix becomes independent of the  of the impurity strength 
in the unitary limit that resembles the vacancies~\cite{SMAH00,SOST06}. 
When replacing $G_0$ by the full Green's function~\cite{SOST06}
\begin{equation}
G(\epsilon,k)=\frac{\epsilon+v_F\tau_3\mathbf{\sigma}\cdot \mathbf{k}}{(\epsilon-\Sigma(\epsilon,k))^2-v_F^2k^2}, 
 \label{eq:green}
\end{equation}
the \textit{selfconsistent  $T$-matrix approximation} is obtained, where $\Sigma$ denotes the particles' 
self-energy in the presence of impurities~\cite{SHU08,SALT06}. 
Performing a disorder-average for $\Sigma$ w.r.t.~the position of the vacancies the self-energy is 
approximated by $\nv$-times the disorder-averaged $T$-matrix. The disorder- average is performed 
separately for the vacancies in the $A$- and in the $B$-sublattice in $k$-space
representation following to Ref.~\cite{SABA10}:
\begin{equation}
\begin{split}
\langle\Sigma(\epsilon)\rangle&=\begin{pmatrix}
        \langle\Sigma_A(\epsilon)\rangle&0\\
0&\langle\Sigma_B(\epsilon)\rangle\\
       \end{pmatrix}\\
       &=\begin{pmatrix}
\nA\langle T_A(\epsilon)\rangle&0\\
0&\nB\langle T_B(\epsilon)\rangle\\\end{pmatrix}.
\end{split}
\end{equation}
Note that $\nA$ and $\nB$ denote the density of impurities w.r.t. the total number of carbon atoms
 in the sample: \[n_{(A/B)}=\frac{N_{(A/B)}}{N_{\text{sites}}}\]
(Here, we employ the convention of the SCTMA-literature where $n_{A/B}$ denotes 
the fraction of $A/B$-vacancies with respect to {\em all} lattice sites. With this convention 
the total concentration of vacancies is given by $\bar n {=} n_A {+} n_B$. ) 

Making use of identity Eq. \eqref{eq:T-Matrix} we derive a set of interdependent equations, 
\begin{equation}
\begin{split}
\langle T\rangle=\begin{pmatrix}
     \langle T_A\rangle& 0\\
0&\langle T_B\rangle\\
    \end{pmatrix}=\frac{1}{ \langle G(\epsilon)\rangle}\\
    \end{split}\end{equation}
    \begin{equation}
    \begin{split}
\langle\Sigma(\epsilon)\rangle=\begin{pmatrix}
\nA\langle T_A(\epsilon)\rangle&0\\
0&\nB\langle T_B(\epsilon)\rangle\\
\end{pmatrix}\\
\end{split}\end{equation}\begin{equation}\begin{split}
\langle G(\epsilon)\rangle&=\int \frac{d^2k}{(2\pi)^2}G(\epsilon,k)\\
&=-\frac{1}{2\pi v_F^2}\log\Big(1-\frac{W^2}{\varepsilon_A\varepsilon_B}\Big)\begin{pmatrix}
\varepsilon_B&0\\
0&\varepsilon_A\\
\end{pmatrix},\\
\end{split}
\label{eq:self_cycle}
\end{equation} which 
require selfconsistent solution~\cite{SABA10}. Here, the abbreviation~\cite{SABA10}
\begin{equation}
\varepsilon_A=\epsilon-\langle \Sigma_A(\epsilon)\rangle;\qquad
\varepsilon_B=\epsilon-\langle \Sigma_B(\epsilon)\rangle
\label{eq:abbreviation}
\end{equation}
has been used; $W\equiv3t$ denotes the bandwidth of the $\pi$-band and 
$v_F$ is the Fermi energy 
We further note that  as shown in 
~\cite{SOST06}, this non-crossing expansion breaks down below an energy scale 
$
\Delta(\nAB) {=} v_\text{F}\sqrt{\pi \nAB/\ln(\nStar/\nAB)   }
$.
As usual,  the density of states can be determined from $\langle G \rangle$ via~\cite{SOST06,SHU08} 
\begin{equation}
\rho(\epsilon)=-\frac{1}{\pi}\Im\Big(\text{tr}(\langle G\big(\epsilon-i0)\rangle\big)\Big).
\label{eq:DOS}
\end{equation}
(When comparing to numerical data from the Krylov space simulation, the SCTMA-result has to be multiplied by 
a factor of $2$ reflecting the existence of two Dirac points in the lattice model.) 

\begin{figure}[htbp!]
\includegraphics[width=0.5\columnwidth]{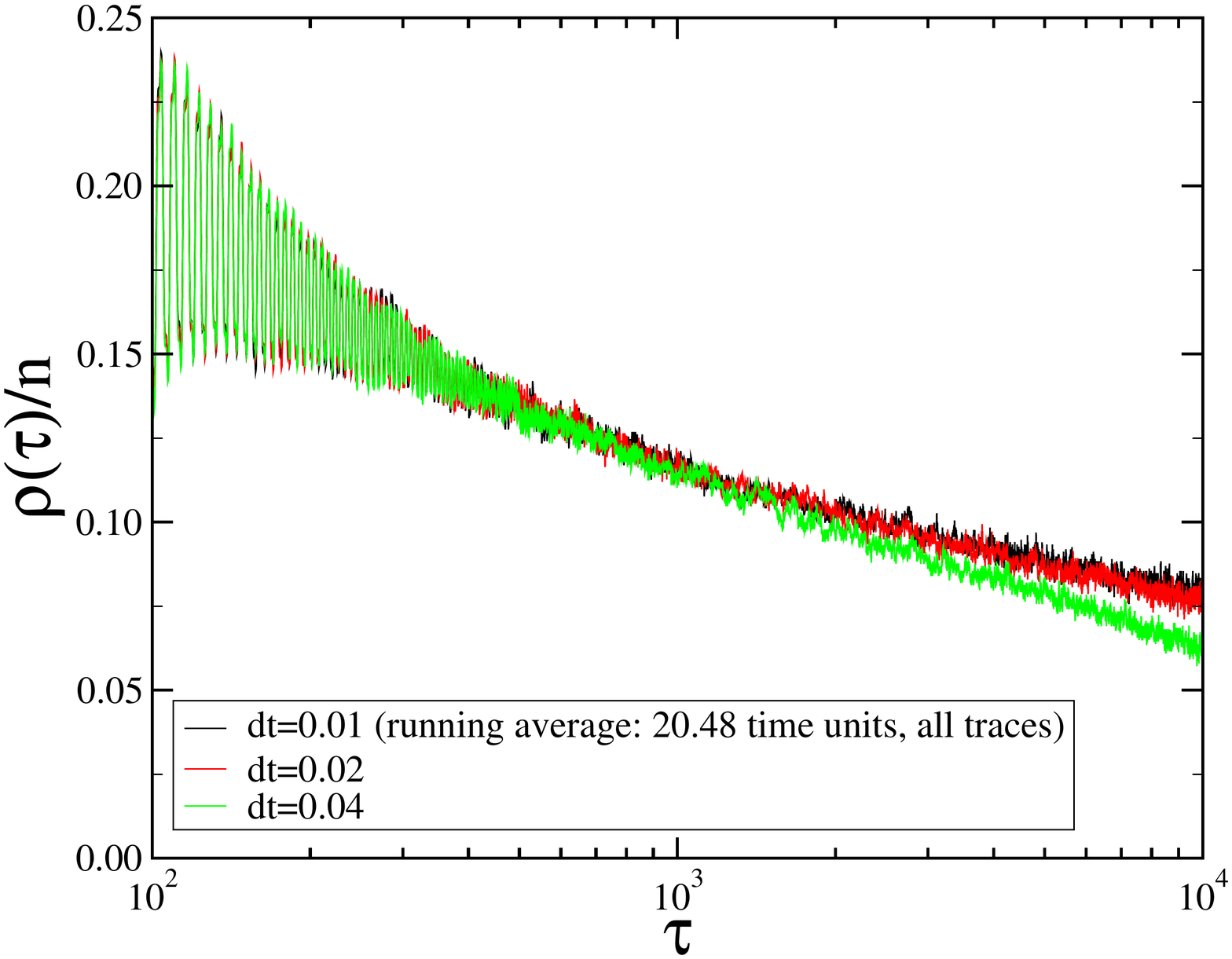}
\caption{\ColorOnline The Fourier-transformed, $\varrho(\tau)$, of the density of states ($\varrho(E)$)
as obtained from the stochastic representation Eq. \eqref{e2} via Krylov-time-propagation for 
an impurity concentration $\nAB=0.5\%$.}
\label{fS1}
\end{figure}
\section{Convergence test for Krylov propagation} 
The Krylov propagation method has two important parameters, the size of the
Krylov space, $N_\text{Krylov}$, and the time increment $dt$. For computational
efficiency one would like to take $dt$ as large as possible while at the same
time keeping $N_\text{Krylov}$ small.  In all our calculations we have chosen
$N_\text{Krylov}=4$.  Fig.~\ref{f2} provides evidence that with this choice a
setting $dt=0.01 t$ is sufficient.  The traces indicate that there is a time
scale associated with each value of $dt$ such that beyond that scale the
correlation function $\varrho(\tau)$ decays too fast.  For instance, with
$dt=0.04$ (Fig.~\ref{fS1}, green) this scale is well below the observation time
$T_\text{obs}=10000$.  On the other hand, the traces for $dt{=}0.01$ (black) and
$dt{=}0.02$ (red) overlap very well within this time window.  The situation is
completely analogous for all other concentrations as well.  For this reason we
consider our choice $dt{=}0.01$ for the time increment as sufficiently
conservative.  

Moreover, we emphasize that computational artifacts related to time propagation
tend to enhance the decay of correlations. Since our numerical calculations
indicate an unexpectedly {\it slow} decay, however, we believe that this aspect
of time propagation is very reliable.

\section{Flow of the IPR-distribution function and freezing} 
\begin{figure}[htbp!]
\includegraphics[width=0.45\columnwidth]{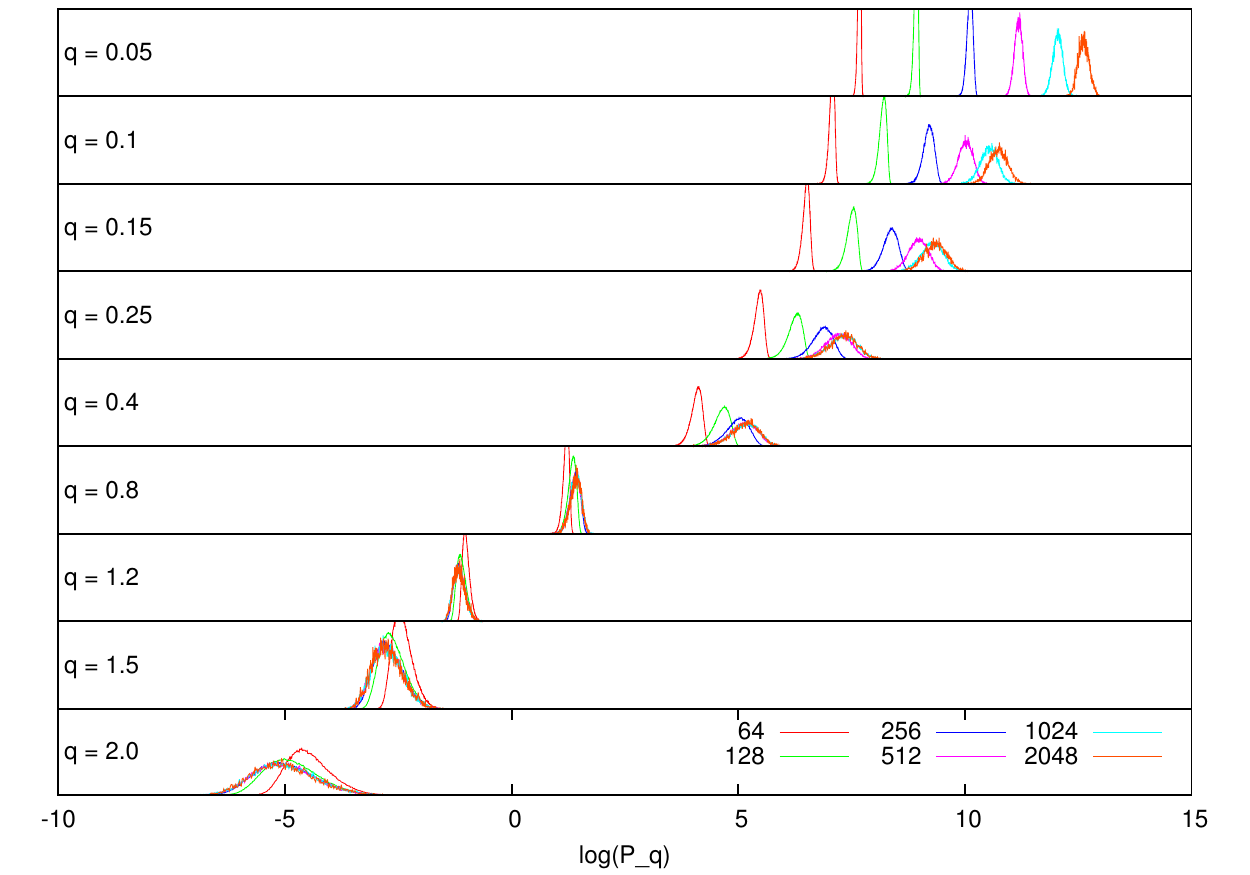}
\includegraphics[width=0.45\columnwidth]{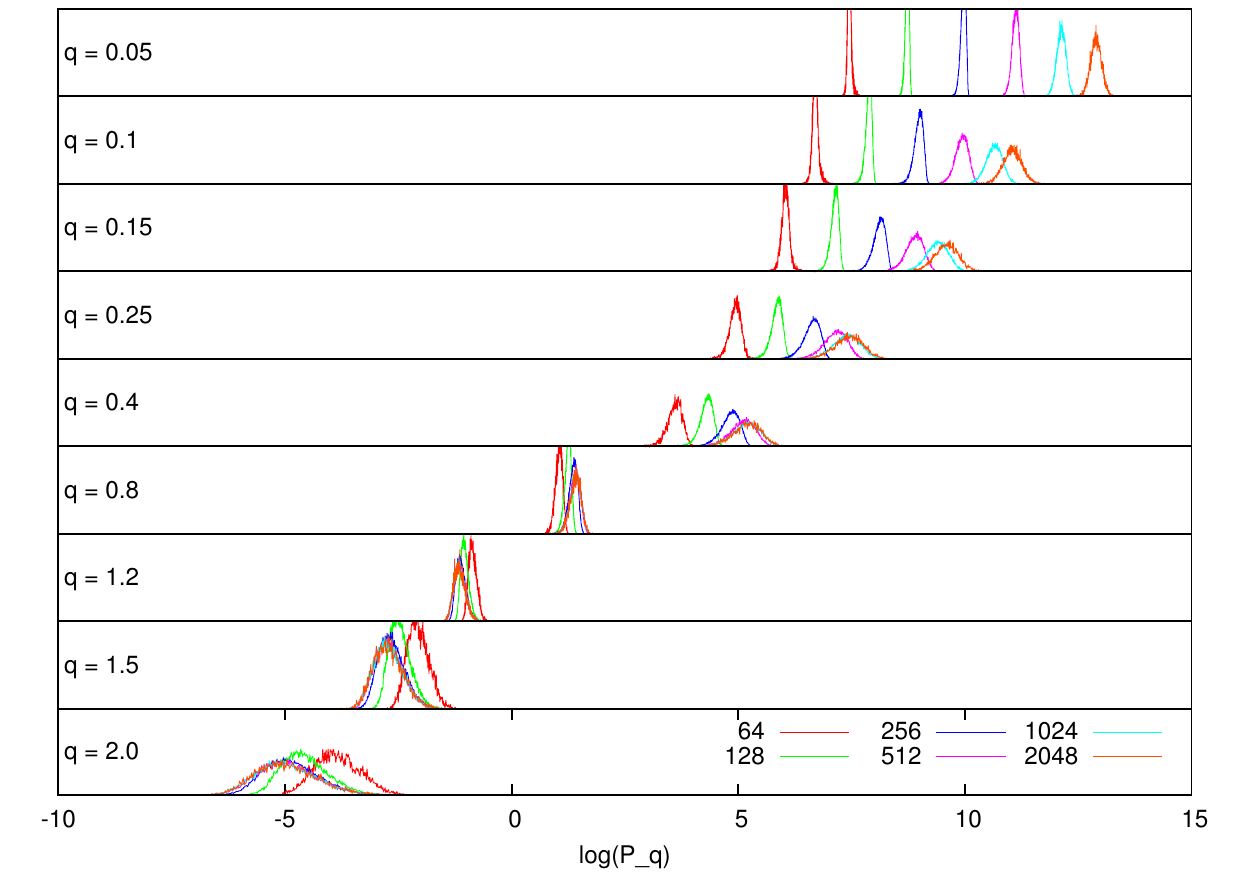}
\caption{\ColorOnline The flow of the distribution function of the inverse participation 
ratio (IPR) for linear system sizes $L=64,128,256,512.1024,2048$ at energies $10^{-3}$
(upper panel) and $10^{-6}$ (lower panel) for selected $q$-values. The 
flow of the average value with system size is captured by the scaling function ${\cal F}(L/\xi(E))$ 
defined in Eq. \eqref{e5} and displayed in Fig.~\ref{f6}. }
\label{fS2}
\end{figure}

At a critical point the distribution function of the (logarithm of the) inverse
participation ratio exhibits a simple scaling property: it flows homogeneously
with increasing system size, $L$, without changing its shape.  As seen in
Fig.~\ref{fS2} there is no such homogeneous flow near the Dirac point. The
behavior is expected at larger system sizes, where conventional localization
sets in so that the distribution function becomes independent of the system size
and the flow stops.  Unexpected is, that the window of system sizes at which
$L\ll \xi(E)$ remains very narrow even when decreasing the observation energies
by three orders of magnitude from $10^{-3}$ down to $10^{-6}t$. By consulting
Fig.~\ref{fS2} convinces oneself that the evolution of the overall flow changes
extremely slowly upon approaching the Dirac point at zero energy consistent with
the almost flat shape of $\xi(E)$ displayed in Fig.~\ref{e6} at ultra-low
energies. 

Remarkably, the strongest changes in the flow are visible at values $q\lesssim
1$.  Here, flow-modifications exist with decreasing the energy in the realm of
small system sizes, e.g., clearly visible at $q=0.15,0.25,0.4$.  By contrast,
there are significantly less modifications at $q>1$ in this regime, comparing
e.g. $q=0.4$ and $q=1.5$.  We take this as preliminary evidence for the presence
of freezing of the IPR which would correspond to $\tau_q{=}0$ at $q>1$ at
strictly zero energy.

\section{Fluctuation effects in the local density of states: exponent $\mfy$} 

We propose a simple heuristic argument indicating that a wide region of energies exist for which 
the typical number of states in the localization volume behaves like $(\sqrt{\nAB}\xi)^{-\mfy}$ with 
$\mfy{=}1$. 
To this end we consider a graphene flake of size $L^2$. We cover it with boxes 
of size $\lambda^2$ where $L\gg \lambda \gg \ell$ and  $\ell$ is a microscopic length. 
Each box contributes on {\em average} a number of states  
\be
Z(\lambda,E) = \lambda^2 \int_{0}^{E} dE' \varrho(E') \sim \lambda^2 E\ \varrho(E).  
\ee 
with energy in the interval $(0,E]$. Suppose that $\epsilon$ is the typical value 
for the smallest energy that a box contributes. 

For weak scatterers (Gaussian disorder) we would expect that $\epsilon$ is of the order 
of the level spacing 
\be
\label{eS15}
\Delta_\lambda = [\lambda^2 \varrho(\Delta_\lambda)]^{-1}
\ee
with fluctuations of order unity. However, vacancies do not appear to fall into this class.

Indeed, consider the fluctuations of the mismatch of the number of vacancies per sublattice 
in each box, $\delta n=\sqrt{\overline{(N_\text{A}-N_\text{B})^2}}$; we have 
$
\delta n^2  \approx 2\nAB \lambda^2.  
$
Now,  a mismatch $\delta n$ is associated with a spectral gap  
$E_\text{gap} \sim \eta v_\text{F} \sqrt{\nAB}$, $\eta=\delta n/\nAB$.  
On SCTMA-level (ignoring possible logarithmic corrections~\cite{weik13}) we get the estimate 
$
E_\text{gap}(\lambda) \sim v_\text{F}/\lambda.
$
Therefore, most boxes exhibit a spectral gap that is much larger than the 
mean level spacing: $E_\text{gap}(\lambda)\gg\Delta_\lambda$. 
Only a small fraction of all boxes, $r(\lambda)$,
can contribute to the total DoS at energies below the gap $E_\text{gap}(\lambda)$. 
We assume, that only those boxes contribute that have a nearly vanishing mismatch. This implies that 
$ r(\lambda)\approx 1/\sqrt{\nAB\lambda^2}$. 
To restore the correct global average,  
the effective DoS in this residual subset of all boxes should be enhanced: 
$\varrho_\text{eff}(E) = \varrho(E)/r(\lambda)$.

\paragraph{Consequences for the localization length.} 
For Gaussian disorder all boxes contribute to the DoS in a similar way. We 
expect a relation for the localization length to the average spectral gap: 
$Z(\xi,\Delta_\xi)\approx {\cal O}(1)$ with $\Delta_\xi \approx \epsilon$. 
As we just have seen, for the case of vacancies the DoS states in 
those boxes that contribute at very low energies is renormalized. 
We extract a localization length from these boxes declaring that  
\be
\xi^2(\epsilon) \epsilon \varrho_\text{eff}(\epsilon) \approx {\cal O}(1) 
\ee
implying  $r(\xi) \sim 1/\sqrt{\nAB\xi^2}$ and $\mfy{=}1$.


\end{document}